\documentstyle[aps,12pt,amssymb,epsfig]{revtex}

\begin{document}

\baselineskip=0.75cm
\newcommand{\ini}{\begin{equation}}
\newcommand{\fin}{\end{equation}}
\newcommand{\inir}{\begin{eqnarray}}
\newcommand{\finr}{\end{eqnarray}}
\newcommand{\inif}{\begin{figure}}
\newcommand{\finf}{\end{figure}}
\newcommand{\bc}{\begin{center}}
\newcommand{\ec}{\end{center}}

\def\ol{\overline}
\def\pa{\partial}
\def\ra{\rightarrow}
\def\ts{\times}
\def\df{\dotfill}
\def\bs{\backslash}
\def\dg{\dagger}

$~$

\hfill DSF-18/2002

\vspace{0.5 cm}

\centerline{{\bf A STATISTICAL MECHANICS FRAMEWORK FOR MULTI-PARTICLE}}

\centerline{{\bf PRODUCTION IN HIGH ENERGY HADRON REACTIONS}}

\vspace{0.5 cm}

\centerline{\large{F. Buccella$^1$ and L. Popova$^2$}}

\vspace{0.5 cm}

\centerline{$^1$Dipartimento di Scienze Fisiche, Universit\`a di Napoli,}
\centerline{INFN, sez. di Napoli}
\centerline{$^2$Inst. of Nucl. Research and Nuclear Energy, Sofia}

\vspace{0.5 cm}

\begin{abstract}

\centerline{Abstract}

$~$
We deduce the particle distributions in particle collisions with
multihadron-production in the framework of mechanical statistics. They are
derived as functions of $x$, $P_T^2$ and the rest mass of different
species for a fixed total number of all produced particles,
inelasticity and total transverse energy.

For $P_T$ larger than the mass of each particle we get the behaviour
\ini
\frac{dn_i}{dP_T} \sim
\sqrt{P_T} e^{-\frac{P_T}{T_H}}
\fin
Values of $<P_T>_\pi$, $<P_T>_K$, and $<P_T>_{\overline{p}}$ in
agreement with experiment are found by taking $T_H=180MeV$ (the Hagedorn
temperature).


\end{abstract}

\newpage

Predictions of multihadron-production beyond accelerator measuraments is
an important tool for study of cosmic rays with high energies. 
Thermodynamical concepts have been advocated for the study of these
processes up to Tevatron energies \cite{cha}. A non-isotropic phase
space should be considered, since most of the energy of the final
state comes from the longitudinal momentum of the final particles.
This has motivated us in a previous work \cite{BP} to look for a
statistical interpretation of the semi-inclusive statistical model
developed by one of us \cite{P}, by approximating the energy of the final
particles
\ini
E_i=\sqrt{P^2_{L_i} + P^2_{T_i} + m^2_i} \simeq |P_{L_i}| +
\frac{P^2_{T_i} + m^2_i}{x_i \sqrt{s}}
\fin
Eq.(2) follows from the assumption for a dominance of the longitudinal
contribution to the total energy and is well verified for
ultra-relativistic particles at small angles with respect to the direction
of the initial particles.

We assume that the missing part of the total longitudinal momentum is
responsible for creation of particles and giving them a transverse
momentum.

Here we perform a mechanical statistical description of this picture by
solving the problem of partition of energy in $N$ produced hadrons,
sharing it in the final longitudinal momentum and the total transverse
energy.

By the classical method of the statistical mechanics one finds for the
distribution of the various particles (in each hemi-sphere):
\ini
dn_i=\frac{1}{M^2}e^{-k x_i}e^{-\frac{P^2_{T_i} + m^2_i}{x_i \mu^2_0}}
\frac{dx_i}{x_i}dP^2_{T_i}
\fin
where $k$ and the dimensional constants $M^2$ and $\mu^2_0$ are
determinated by the boundary conditions:
\ini
\sum_i \int{dn_i} = N
\fin
\ini
\sum_i \int{x_i dn_i} = \eta_{inel} - \frac{2E_T}{\sqrt{s}}
\fin
\ini
\sum_i \int{ \frac{P^2_{T_i} + m^2_i}{x_i \sqrt{s}} dn_i} = E_T
\fin
Here the sums exclude the leading particle and $\eta_{inel}$ and $E_T$ are
the inelasticity and the total transverse energy, respectively (the
latter has a negligible contribution with respect to the longitudinal one
in Eq.(5)).

From Eq.(3) it is easy to derive the total number of particles of the
species $i$:
\ini
n_i = \frac{\mu^2_0}{M^2} \int{dx e^{-kx} e^{-\frac{m^2_i}{x \mu^2_0}}}
\fin
by keeping into account that the function to be integrated has a
sharp maximum around the minimum, $x^{min}_i=m_i/\mu_0\sqrt{k}$, of
\ini
f_i(x) = kx + \frac{m^2_i}{x \mu^2_0}
\fin
and $f_i(x)$ may be well approximated by
\ini
f_i(x^{min}_i) + \frac{1}{2} f''_i(x^{min}_i) (x - x^{min}_i)^2
\fin
which gives rise to a gaussian integral.

In Table I we report $x^{min}_i$, $f_i(x^{min}_i)$ and $f''_i(x^{min}_i)$
and the corresponding values for the functions
\ini
g^{p}_i(x)=f_i(x) -p \ln(x)
\fin
which appear in the expression of $<P_{T_i}>$, $<P^2_{T_i}>=\mu^2_0 <x_i>$
and of the contribution of the term proportional to $m^2_i$ to the
total transverse energy in correspondence of $p=1/2$, $1$ and $-1$,
respectively.

From Table I it is easy to derive  within the gaussian
approximation for the integrals in the variable $x$:
\ini
n_i =
2\sqrt{2\pi}\frac{T^2_H}{M^2}e^{-\frac{m_i}{T_H}}\sqrt{\frac{m_i}{T_H}}
\fin
\ini
n_i<P_{T_i}> = 2\pi \frac{T^3_H}{M^2}
\left( \frac{1}{2} + \sqrt{\frac{1}{4}+\frac{m^2_i}{T^2_H}} \right)^{3/2}
\frac{e^{-\sqrt{\frac{1}{4} + \frac{m^2_i}{T^2_H}}}}{\left( \frac{1}{4} +
\frac{m^2_i}{T^2_H} \right)^{1/4}}
\fin
\ini
n_i <P^2_{T_i}> = n_i \mu^2_0 <x_i> =
4 \sqrt{2\pi} \frac{T^4_H}{M^2} \left( 1 + \sqrt{1+\frac{m^2_i}{T^2_H}}
\right)^{2}
\frac{e^{-\sqrt{1 + \frac{m^2_i}{T^2_H}}}}{\left(1 +
\frac{m^2_i}{T^2_H}\right)^{1/4}}
\fin
\ini
n_i<\frac{P^2_{T_i}}{x\sqrt{s}}>=n_i \frac{\mu^2_0}{\sqrt{s}}
\fin
\ini
n_i<\frac{m^2_i}{x\sqrt{s}}> = \sqrt{2\pi}
\frac{\mu^2_0 m^2_i}{M^2\sqrt{s}}
\frac{e^{-\sqrt{1+\frac{m^2_i}{T^2_H}}}}{\left(1 
+\frac{m^2_i}{T^2_H}\right)^{1/4}}
\fin
where
\ini
T_H=\frac{\mu_0}{2\sqrt{k}}
\fin
The expression in eq.(3) gives the correlation known as "sea-gul" effect
(with the square of the transverse momentum increasing with $x$). It also
implies a larger transverse momentum for the heavier particles, which are
produced at a larger $x$ due to the presence of the exponential factor
$exp(-m^2_i/x\mu^2_0)$.

The ratio
\ini
\frac{<P^2_{T_i}>}{<x>_i} = \mu^2_0
\fin
does not depend on the species and this property is an important check of
this statistical approach. By integrating eq.(3) in $x$, again with the 
gaussian approximation for the function, one obtains from $g^{-1}(x)$
by the substitution $m^2_i \rightarrow m^2_i +P^2_{T_i}$, one gets the
$P_T$ distributions:
\ini
\frac{dn_i}{dP_{T_i}}=
\frac{2\sqrt{2\pi}P_{T_i}}{M^2}
\frac{e^{-\sqrt{1+\frac{m^2_i + P^2_{T_i}}{T^2_H}}}}{\left(1 + \\
\frac{m^2_i+P^2_{T_i}}{T^2_H}\right)^{1/4}}
\fin
which, for $P_{T_i} \gg m_i$, $T_H$ approaches:
\ini
\frac{dn_i}{dP_{T_i}}=\frac{2\sqrt{2\pi P_{T_i} T_H}}{M^2}
e^{-\frac{P_{T_i}}{T_H}}
\fin
So we recover the exponential behaviour found experimentally. By fixing
$T_H=180MeV$ (the Hagedorn temperature \cite{H}), we predict $<P_{T_i}>$
as a function of $m_i$ and $T_H$:
\ini
<P_{T_i}>=\frac{ \sqrt{\frac{\pi}{2}} T_H 
\left( \frac{1}{2}+ \sqrt{\frac{1}{4}+\frac{m^2_i}{T^2_H}} \right)^{3/2}
e^{-\frac{1}{4 \left(\frac{m_i}{T_H}+ \sqrt{\frac{1}{4}+
\frac{m^2_i}{T^2_H}}\right) } } }
{\sqrt{\frac{m_i}{T_H}} \left(\frac{1}{4}+
\frac{m^2_i}{T^2_H}\right)^{1/4}}
\fin
which gives:
\ini
<P_T>_\pi = 390 MeV
\fin
\ini
<P_T>_K = 467 MeV
\fin
\ini
<P_T>_{\overline{p}} = 580 MeV
\fin
to be compared with the experimental results at $s=(540GeV)^2$, $372MeV$,
$482MeV$ and $606MeV$ respectively \cite{dat}.
Also we find:
\ini
\frac{n_{K^\pm}}{n_{\pi^\pm}} = \sqrt{\frac{m_K}{m_\pi}}
e^{\frac{m_\pi-m_K}{T_H}}=0.256
\fin
\ini
\frac{n_{\overline{p}}}{n_{\pi^-}} = 2\sqrt{\frac{m_N}{m_\pi}}
e^{\frac{m_\pi-m_N}{T_H}} = 0.061
\fin
while the experimental values are $0.112\pm0.010$ and $0.078\pm0.011$,
respectively \cite{dat}. The approximation described
in eq.(2) is valid as long the $P_T$ distribution is considered for:
\ini
8k^2 \left( m^2_i+P^2_{T_i} \right) \ll s \left( \sqrt{1+ \frac{m^2_i
+P^2_{T_i}}{T^2_H}} -1 \right)^2
\fin
which is very well satisfied for any value of $m_i$ and $P_{T_i}$, if
one takes $k=16.5$, as it follows from the value $7$ found for $k$ by
\cite{P} at $s=(20GeV)^2$ by assuming:
\ini
k(s)=k(s_0) \left( \frac{s}{s_0}\right)^{0.13}
\fin
The high value of the second derivatives in Table 1 support the validity
of the gaussian approximation. Also the functions $f_i(x)$ and
$g^{ip}(x)$ are not so different, since they differ by a term
proporional to $\ln(x)$, and so we expect a better precision for the
ratios of the quantities described in eqs.(11-13),which give $<P_{T_i}>$,
$<P^2_{T_i}>$ and $<x_i>$.
From eq.(16) with $T_H=180GeV$ and $k=16.5$, one finds at
$\sqrt{s}=540GeV$, $\mu_0 \simeq 1.46GeV$, which implies for the average
contribution to the transverse energy:
\ini
<\frac{P^2_{T_i}}{x\sqrt{s}}>=\frac{\mu^2_0}{\sqrt{s}} \simeq
4MeV
\fin
for the transverse momentum (independent of the species) and
\ini
<\frac{m^2_i}{x\sqrt{s}}>=\frac{\mu^2_0}{2\sqrt{s}}
\frac{m_i e^{-\frac{T_H}{m_i+\sqrt{T^2_H+m^2_i}}}}{T_H
\left(\frac{T^2_H}{m^2_i}+1\right)^{1/4}}
\fin
which gives $0.56MeV$, $4.4MeV$ and $18.4MeV$ for $\pi$, $K$ and
$\overline{N}$, respectively.

The good results found for $P_T$ distributions support the statistical
approach described here. The large value found for $n_K/n_\pi$
and slightly small for $n_{\overline{p}}/n_\pi$ may be related to have not
considered the fluctuations in the r.h.s.'s of eqs.(4-6).
In particular we know that the number of particles produced at a given
energy is not even a gaussian (or a Poissonian), as realized from
the distribution of the relative multiplicity $n/<n>$ found many years ago
\cite{P} and being also analysed in \cite{chyang}. In fact we may expect
that the tail at large $n$ will favour a copious pion production, while
for small $n$ antiprotons will be less disfavoured by the factor
$e^{-m^2_i/x\mu^2_0}$ by the larger $x$ involved.
For the purpose of cosmic ray study we have to find the energy dependence
of the quantities we have fixed ($n$, $\eta_{inel}$ and $E_T$) or
of the associated Lagrangian multipliers ($k$, $\mu$) and $M^2$, the
dimensional constant related to $n$.
For $n$ we may assume the power behaviour $(s/s_0)^\alpha$ with
$\alpha=0.13$ succesfully assumed in the semi-inclusive statistical model
\cite{P} and confirmed by the empirical expression found at Tevatron
experiments $(=-7+7.2 (s/s_0)^{0.127\pm0.005})$ \cite{Tev}. For $k$ we
could assume, as we have already done in checking the validity of eq.(2), 
the same behaviour implied by the use of the variable
$x_s=x(s/s_0)^\alpha$ \cite{ww}. Finally the energy dependence of
$T_H=\frac{\mu_0}{2\sqrt{k}}$ may be found by studying the high $P_T$
behaviour of $dn_i/dP_T$ as given by eq.(20), as a function of energy:
the increase with $s$ of the mean transverse momenta implies a
corresponding behaviour \cite{dat} of $T_H$ according to eq.(20). A
realistic description will require the consideration of the fluctuations.

The success of the statistical approach developed for the production of
strange particles in $e^+ e^-$ reactions \cite{BEC} makes us
confident in applying the approach described here to the production
of the particles composed of the light $SU(3)$ quarks $(u,~d,~s)$. However
the event found in the associated charm production in deep-inelastic
neutrino induced reactions at CHORUS \cite{CHORUS} may be a signal for a
larger production than the evaluations based
on elementary $QCD$ processes with amplitudes proportional to $\alpha_S$
and motivate the extension to the charmed particles of the
statistical approach described here.

\newpage

\begin{center}
\begin{tabular}{|c|c|c|}
\hline
$x^{i}_{min}=\frac{m_i}{\mu_0\sqrt{k}}$  &
$f_i(x^i_{min})=2\frac{\sqrt{k}}{\mu_0}m_i$ &
$f''_i(x^i_{min})=\frac{2 \mu_0 k^{3/2}}{m_i}$
\\ \hline
$x^{i;p}_{min}=\frac{p+\sqrt{p^2 +
\frac{4km^2_i}{\mu^2_0}}}{2k}$ &
$g^p_i(x^{i;p}_{min})=\sqrt{p^2+\frac{4km^2_i}{\mu^2_0}}-
p\ln x^{i;p}_{min}$ &
$g^{p}{''}_i(x^{i;p}_{min})=
\frac{\sqrt{p^2+\frac{4km^2_i}{\mu^2_0}}}{(x^{i;p}_{min})^2} $
\\ \hline
\end{tabular}
\bigskip \\[0pt]
Table I
\end{center}


\begin{references}

\newpage

\centerline{\large{References}}

\baselineskip=0.6cm

\bibitem{cha} T.T.Chou and C.N.Yang,
Phys.Rev. D 32, (1985) 1692

\bibitem{BP} F.Buccella and L.Popova,
Nuovo Cimento A 112, (1999) 253

\bibitem{P} L.Popova,
J. Phys. G 9, (1983) 243

\bibitem{H} R.Hagedorn
Riv. Nuovo Cimento 10, (1983) 1

\bibitem{dat} T.Alexopoulos et al (E735 coll.),
Phys. Rev. D 48, (1993) 984

\bibitem{chyang} T.T.Chou and C.N.Yang,
Phys. Rev. Lett 54, (1985) 510

\bibitem{Tev} F.Abe et al (CDF coll.),
Phys. Rev. D 50, (1994) 5550

\bibitem{ww} J.Wdowczyk and A.Wolfendale,
Nuovo Cimento A 54, (1979) 433

\bibitem{BEC} F.Becattini,
World Scientific "Universality features in multihadron
production and leadingeffect", (1998) 74

\bibitem{CHORUS} A.~Kayis-Topaksu {\it et al.} [CHORUS Collaboration],
Phys.\ Lett.\ B {\bf 539}, (2002) 188

\end{references}
\end{document}